\documentclass[12pt]{article}

\newtheorem{theorem}{Theorem}[section]

\newtheorem{defn}{Definition}[section]

\newcommand{\beqn}{\begin{equation}}
\newcommand{\eeqn}{\end{equation}}
\newcommand{\be}{\begin{eqnarray}}
\newcommand{\ee}{\end{eqnarray}}
\newcommand{\bes}{\begin{eqnarray*}}
\newcommand{\ees}{\end{eqnarray*}}
\newcommand{\bmini}{\begin{minipage}}
\newcommand{\emini}{\end{minipage}}

\newcommand{\Pa}{Painlev\'e}
\newcommand{\Complex}{\textrm{\kern.24em \vrule
     width.02em height1.2ex
     depth-.05ex\kern-.26em C}}
\newcommand{\Real}{\textrm{I\kern-.20em R}}
\newcommand{\li}[1]{\begin{list}{\arabic{enumi}.}
                     {\usecounter{enumi}
		                    \setlength{\leftmargin}{8truept}
                      \setlength{\itemsep}{1pt}
                      \setlength{\parsep}{0pt}
                      }{{#1}}\end{list}}

\newcommand{\A}{\alpha}
\newcommand{\B}{\beta}
\newcommand{\C}{\gamma}
\newcommand{\D}{\delta}
\newcommand{\E}{\epsilon}

\newcommand{\LA}{\lambda}

\newcommand{\T}{\tau}

\newcommand{\half}{{{1}\over{2}}}
\newcommand{\nn}{\nonumber}
\newcommand{\rf}[1]{(\ref{#1})}

\begin{document}
\title{Analytic and Asymptotic Methods for
Nonlinear Singularity Analysis: a Review and Extensions of Tests
for the Painlev\'e Property}
\author{
\textit{Martin D.~Kruskal}$^1$, \textit{Nalini Joshi
}$^{2}$, and \textit{Rod Halburd}$^3$}
\footnotetext[1]{
 Department of Mathematics,
	Rutgers University,
	New Brunswick NJ 08903
	USA, \texttt{kruskal@math.rutgers.edu}
 }
\footnotetext[2]{
 Department of Pure Mathematics,
	University of Adelaide,
	Adelaide SA 5005
	Australia, \texttt{njoshi@maths.adelaide.edu.au} }
\footnotetext[3]{
 Department of Applied Mathematics,
	University of Colorado,
	Boulder CO 80309-529
	USA \texttt{rod@newton.colorado.edu}
}
\renewcommand{\thefootnote}{\fnsymbol{footnote}}
\maketitle
\begin{abstract}
The integrability (solvability via an associated single-valued
linear problem) of a
differential equation is closely related to the singularity structure
of its solutions.  In particular, there is strong evidence that
all integrable equations have the Painlev\'e property, that is, all
solutions are single-valued around all movable singularities.
In this expository article, we review methods for analysing
such singularity structure. In particular, we describe well known
techniques of nonlinear regular-singular-type analysis, i.e.
the Painlev\'e tests for ordinary and partial differential equations.
Then we discuss methods of obtaining sufficiency conditions for the
\Pa\ property. Recently, extensions of
\textit{irregular} singularity analysis to nonlinear equations
have been achieved. Also, new asymptotic limits of differential equations
preserving the \Pa\ property have been found. We discuss
these also.
\end{abstract}
\section{Introduction}
A differential equation is said to be integrable if it is
solvable (for a sufficiently large class
of initial data) via an associated (single-valued)
linear problem.  A famous example
is the
Korteweg-de~Vries equation (KdV),
\begin{equation}
        u_t+6uu_x+u_{xxx}=0,    \label{kdv}
\end{equation}
where the subscripts denote partial differentiation.

The KdV equation was discovered to be integrable by
Gardner, Greene, Kruskal, and Miura~\cite{gardnergkm}.
(Its method of solution is called the \textit{inverse scattering
transform} (IST) method; see the paper by  Mark Ablowitz in
the present collection.) Since this discovery, a large collection
of nonlinear equations (see \cite{ablowitzc}) has been identified to be
integrable. These range over many dimensions and include
not just partial differential equations (PDEs) but also
differential-difference
equations, integro-differential equations, and ordinary differential
equations (ODEs).

Six classical nonlinear second-order ODEs called the \Pa\ equations
are prototypical examples of integrable ODEs.
They possess a characteristic singularity structure
i.e. all movable singularities of all
solutions are poles. Movable here means that the singularity's
position varies as a function of initial
values. A differential equation is said to have the
\Pa\ property if all solutions are single-valued around all
movable singularities. (See comments below and in Section 2 on variations
of  this definition.) Thefore, the \Pa\ equations possess
the \Pa\ property.
\Pa\ \cite{painleve02}, Gambier \cite{gambier}, and
R. Fuchs \cite{fuchs} identified these equations (under some
mild conditions) as the only ones (of second order and first
degree) with the Painlev\'e property whose general solutions
are new transcendental functions.

Integrable equations are rare. Perturbation of such equations
generally destroys their integrability. On the other hand,
any constructive method of identifying the integrability of a
given system contains severe shortcomings. The problem is
that if a suitable associated linear problem cannot be found
it is unclear whether the fault lies with the lack of
integrability of the nonlinear system or with the lack of
ingenuity of the investigator. So the identification of
integrability has come to rely on other evidence, such as
numerical studies and the singularity structure of the system.

There is strong evidence \cite{yoshida,ziglin} that
the integrability of a nonlinear sytem is intimately related
to the singularity structure admitted  by the system in its
solutions.  Dense multi-valuedness (branching) around movable
singularities of solutions is an indicator of
nonintegrability\cite{ziglin82}.  The \Pa\ property excludes such branching
and has been proposed as a pointer to integrability.

The complex singularity structure of solutions was first used
by  Kowalevs-kaya \cite{kowalevskayaa,kowalevskayab} to
identify an integrable case of the equations of motion for a
rotating top.  Eighty eight years later, this connection was
reobserved in the context of integrable PDEs by Ablowitz and
Segur~\cite{ablowitzs}, and Ablowitz, Ramani, and
Segur~\cite{ablowitzrs1,ablowitzrs2}. Their observations led
to the following conjecture.
\begin{quotation}
{\bf The ARS Conjecture:} {\em
Any ODE which arises as a reduction of an integrable PDE
possesses the Painlev\'e property, possibly after a
transformation of variables.}
\end{quotation}

For example, the sine-Gordon equation
\begin{equation}
	u_{xt}=\sin u,	\label{sinegordon}
\end{equation}
which is well known to be integrable \cite{ablowitzsegur,ablowitzc},
admits the simple scaling symmetry
\begin{equation}
	x\to \LA x,\quad t\to\LA^{-1}t.	\label{sgsym}
\end{equation}
To
find a reduction with respect to the symmetry \rf{sgsym}, restrict to the
subspace of solutions that is invariant under
\rf{sgsym} by introducing new variables $z$, $w$ such that
$$ u(x,t)=w(z),\quad z=xt.	$$
This gives
\begin{equation}
		zw''+w'=\sin w,	\label{prepiii}
\end{equation}
where the prime denotes differentiation with respect to $z$.
To investigate the Painlev\'e property, this equation must
first be transformed to one that is rational (or possibly
algebraic) in $w$. (Otherwise, the nonlinear analogue of
Fr\"obenius analysis  used to investigate the \Pa\ property
cannot find a leading-order term to get started.
See Section 2.)
Introduce the new dependent variable $y:=\exp (iw)$. Then
equation~\rf{prepiii} becomes
$$	z(yy''-y'^2)+yy'=\half y(y^2-1).	$$
This equation (a special case of the third Painlev\'e
equation) can be shown to have the Painlev\'e property. (See
Section 2.)

There is now
an overwhelming body of evidence for the ARS conjecture. A
version that is directly applicable to PDEs, rather than
their reductions, was given by Weiss, Tabor, and Carnevale
(WTC)\cite{weisstc}. The ARS conjecture and its variant by
WTC are now taken to be almost self-evident because they have
been formally verified for every known analytic soliton
equation \cite{kruskalc,ramanigb,newelltz}  (where analytic
means that the equation is, or may be written to be, locally
analytic in the dependent variable and its derivatives).
Previously unknown integrable versions of the soliton
equations \cite{joshi87,clarksonc,hlavaty} have been
identified by the use of the conjecture. The conjecture has
also been extrapolated to identify integrable ODEs
\cite{bountissv,ramanidg}.

Rigorous results supporting the conjecture exist for ODEs with
special symmetries or symplectic structure
\cite{yoshida,ziglin}. Necessary conditions for possessing
the \Pa\ property
\cite{cosgrove,hlavaty90} have also
been derived for general semilinear analytic second-order PDEs.
No new integrable
PDEs were found --- suggesting that in this class at least,
there is a one-to-one correspondence (modulo allowable
transformations) between integrable equations and those
possessing the \Pa\ property. Moreover, proofs of weakened
versions of the conjecture exist \cite{ablowitzs,mcleodo}.
These results point strongly to the truth of the ARS
conjecture. Nevertheless, the conjecture has not yet been
proved.

The main aim of this paper is to describe methods
for investigating the singularity structure of
solutions of ODEs and PDEs. These may be divided
into two classes, those that parallel methods for
analysing \textit{regular} singular points
and those that parallel techniques for \textit{irregular} singular
points of linear ODEs.

The first class of methods has been widely used formally.
The most popular procedure is to expand every solution of
the differential equation of interest in an infinite series
near a movable singularity of the equation \cite{kruskalj},
i.e. the solution $u(z)$ is expanded as
\beqn
u(z)=\sum_{n=0}^{\infty}a_n(z-z_0)^{n+\rho}
\label{pseries}
\eeqn
where $z_0$ is the arbitrary location of a singularity
and $\rho$ is the leading power that needs to be
found. Such an expansion is often called a \Pa\ expansion.
The equation is assumed to have the \Pa\ property if
the series is self-consistent, single-valued,
and contains a sufficient number of
degrees of freedom to describe all possible solutions
or the general solution. These demands yield
necessary conditions for the \Pa\ property to hold. The series
and the expansion techniques are analogues of the usual
Fr\"obenius (or Fuchsian)  expansion procedure for linear
ODEs. This procedure was extended to PDEs by \cite{weisstc}.

These techniques are, in general, not sufficient to prove
that a differential equation has the \Pa\ property. For
example, even if the only possible formal solutions are Laurent
series, the poles indicated by these series may accumulate
elsewhere to give rise to a worse (branched) singularity.

\Pa\ gave sufficient conditions to show that his eponymous
equations have the \Pa\ property. However, his proof is not
widely understood. We describe briefly here an alternative,
direct, method of proof due to Joshi and Kruskal
\cite{joshik94}. To gain sufficiency, we showed that the
solutions of the \Pa\ equations possess convergent Laurent
expansions around every movable singularity, and moreover,
(in any given bounded region) the radius of convergence of
each series is uniformly bounded below. In other words, the
poles of any solution cannot coalesce to form a more
complicated singularity elsewhere (in the finite plane).

For nonlinear PDEs, the question of how to get
sufficient conditions for the \Pa\ property is still open.
Nevertheless, partial results are now known. These make
the WTC analogue of the Fr\"obenius method rigorous and go
some way toward proving that a given PDE has the \Pa\
property. We describe the results due to Joshi and Petersen
\cite{joship1,joship2} and Joshi and Srinivasan \cite{joshis}
in section 2. An alternative approach to the convergence of the
\Pa\ expansions for PDEs has
also been developed by
Kichenassamy and Littman
\cite{kichenassamyl,kichenassamys}.

Another difficulty with the analysis of singularity
structure is that the Painlev\'e
expansions can miss some solutions. This
may happen, for example, when the number of degrees of
freedom in the series is less than the order of the
differential equation. Perturbations of such series often reveal
that the missing degrees of freedom lie in terms that
occur (paradoxically) before the leading term.
For reasons explained in Section 2, such terms are called
\textit{negative resonances}.
In other cases, perturbations reveal no additional degrees
of freedom at all.
We call the latter series
\textit{defective}.

How can we deduce
the singular behaviours of solutions that are missed
by the \Pa\ expansions?  We provide an answer based on irregular-singular
analysis for linear ODEs \cite{bendero} and illustrate it through two
important examples.
The first example is the Chazy
equation \cite{ablowitzc}, a third-order ODE, whose general solutions  have
movable natural barriers. The \Pa\ expansion of the solution of the
Chazy equation contains only two arbitrary constants. The second
example is a fourth-order ODE first studied by
Bureau.
This example has two families of \Pa\ expansions,
one of which has negative resonances and one that is defective.
In section 3, we show how the \Pa\ expansions can be extended through
exponential (or WKB-type) perturbations.

Conte, Fordy, and Pickering \cite{contefp} have followed an
alternative approach. Their perturbations of \Pa\ expansions involve
Laurent series with no leading term (i.e. an infinite number of
negative powers). As pointed out by one of us, this is well defined
only in an annulus where the expansion variable is lower-bounded
away from the singularity. Conte \textit{et al.} demand that
each term of such a perturbation must be single-valued. Therefore,
their procedure requires a possibly infinite number of conditions
to be checked for the \Pa\ property. Our approach overcomes this
problem.

For linear differential equations, in general, the analysis
near an irregular singularity yields asymptotic results, i.e.
asymptotic behaviours along with their domains of asymptotic validity
near the singularity. The latter is crucial in this description.
For example, it is well known that the Airy function $Ai(x)$
which solves the ODE
\[
y''=xy,
\]
has the asymptotic behaviour
\[
Ai(x)\approx{1\over 2\sqrt\pi x^{1/4}}\exp\bigl(-2x^{3/2}/3\bigr)
\quad\textrm{as }\ |x|\to\infty,\quad |\arg x|<\pi
\]
near the irregular singular point at infinity.
(See \cite{bendero,olver:special}.) Note that the asymptotic
behaviour of $Ai(x)$ is apparently multivalued but the function
itself is single-valued everywhere. (In fact, $Ai(x)$
is entire, i.e. it is analytic throughout the whole  complex
$x$-plane.)

The resolution of this
apparent paradox lies in the angular width of the sector of
validity of the above behaviour, which is strictly less than
$2\pi$. To describe $Ai(x)$ in the whole plane, we need
its asymptotic behaviour in regions that include the line
$|\arg(x)|=\pi$. (Such behaviours are well known. See e.g.
\cite{bendero}.) These, together with the behaviour given
above, show that the analytic continuation of $Ai(x)$ along a
large closed curve around infinity is single-valued.
Therefore, the global asymptotic description is not actually
multivalued.

On the other hand, suppose an asymptotic behaviour is
multivalued and its sector of validity extends further than
$2\pi$. Then there are (at least) two asymptotic descriptions
of a solution at the same place (near an irregular
singularity). This violates the uniqueness of asymptotic
description of a solution, unless the solution is itself
multivalued. Therefore, such an asymptotic behaviour is an
indication that the solution cannot satisfy the requirements
of the \Pa\ property.  The Bureau equation we study in
Section 3 provides an example of this case.

Such results form an important extension of
the usual tests for the \Pa\ property. However, there is no
denying that many fundamental questions remain open
in this area, even at a formal level. For example, the \Pa\
property is easily destroyed by straightforward
transformations of the dependent variable(s). (E.g. A solution
$u(z)$ of an ODE with movable simple poles is transformed to
a function $w(z)$  with movable branch points under $u\mapsto
w^2$.) An extension of the \Pa\ property called the poly-\Pa\
property has been proposed by Kruskal \cite{kruskalc,ramanigb}
to overcome these difficulties.
It allows solutions to be
branched around movable singularities so long as
a solution is not densely valued at a point. However, such
developments lie outside the scope of this paper and we refer
the reader to
\cite{kruskalc} for further details and references.

Other
major problems remain.
One is to extend the classification work of \Pa\ and his
colleagues to other classes of differential equations. Cosgrove
has accomplished the most comprehensive extensions in recent
times \cite{cosgrove}. The universal method of classification
called the $\alpha$-method is based on asymptotic
ideas (see Section 2).
Asymptotic limits of differential equations can illuminate such
studies.

The \Pa\ equations are well known to have asymptotic limits
to other equations with the \Pa\ property.
These limits are called
\textit{coalescence} limits because singularities of the equation
merge under the limit. In Section 4, we
describe the well known coalescence limits of the \Pa\ equations and show that
these limits also occur for integrable PDEs.

Throughout this article, solutions of
differential equations assumed to be complex-valued
functions of complex variables.

\section{Nonlinear-Regular-Singular Analysis}
In this
section, we survey the main techniques used to
study the \Pa\ property. These range from the $\alpha$-method
to the widely used formal test known as the \Pa\ test.

Consider the second-order linear ODE
$$
u''(z)+p(z)u'(z)+q(z)u(z)=0,
$$
where primes denote differentiation with respect to $z$.
Fuchs' theorem \cite{bendero} states that $u$ can only be
singular (nonanalytic) at points where $p$ and $q$ are singular.
Such singularities are called {\em fixed} because
their positions are determined \textit{\`a priori} (before
solving the equation) and their locations
remain unchanged throughout the space of all possible
solutions.

However, fixed singularities are not the only possibilities
for
nonlinear equations. Consider the
Riccati equation
$$
u''(z)+u^2(z)=0.
$$
It has the general solution
$$
u(z)={1\over z-z_0},
$$
where $z_0$ is an arbitrary constant.
If, for example, the initial condition
is $u(0)=1$, then $z_0=-1$. If the initial
condition is changed to $u(0)=2$, then $z_0$ moves to
$z_0=-1/2$. In other words, the location of the singularity
at $z_0$
\textit{moves} with initial conditions. Such singularities
are called \textit{movable}.

Nonlinear equations exhibit a vast range of types of
movable singularities.  Some examples are given in the
Table~1 (where $k$ and $z_0$ are arbitrary constant
parameters).
\begin{center}
\begin{tabular}{|c|p{3.0cm}|c|p{3cm}|}
\hline
\multicolumn{4}{|c|}{\bf Table 1: Examples of Possible Singular Behaviour}\\
\hline
&Equation & General Solution & Singularity Type\\
\hline\hline
&&&\\
1.& $y'+y^2=0$&
$ y={(z-z_0)^{-1}}      $ & \bmini{3cm}simple pole\emini\\
&&&\\
\hline
&&&\\
2.&$ 2yy'=1 $ &
$ y=\sqrt{z-z_0} $    &  \bmini{3cm} branch point\emini \\
&&&\\
\hline
&&&\\
3.&$ y''+y'^2=0 $ &
$y=\ln (z-z_0)+k  $ &
\bmini{3cm} \begin{center}logarithmic\\branch point\end{center}\emini\\
&&&\\
\hline
&&&\\
4.&\begin{minipage}{3.0cm}
\begin{eqnarray*}& &yy''\\
  &+&y'^2(y/y'-1)\\
  &=&0
\end{eqnarray*}\end{minipage}&
$ y=k\exp\left(
        [z-z_0]^{-1}\right) $ &
\begin{minipage}{3.0cm}\begin{center} isolated\\
essential\\singularity\end{center}
\end{minipage}\\
&&&\\
\hline
&&&\\
5.&\begin{minipage}{3.0cm}
\begin{eqnarray*}
& &(1+y^2)y''\\&+&(1-2y)y'^2
\\&=&0
\end{eqnarray*}\end{minipage}&
    $ y=\tan\left(\ln(k[z-z_0])\right)$ &
\begin{minipage}{3.0cm}\begin{center} nonisolated\\essential\\
singularity\end{center}
\end{minipage}\\
&&&\\
\hline
&&&\\
6.&\begin{minipage}{3.0cm}
\begin{eqnarray*}& &\bigl(y''+y^3y'\bigr)^2\\
&=&y^2y'^2\bigl(4y'+y^4\bigr)
\end{eqnarray*}\end{minipage}&
$y=k\tan\bigl[k^3(x-x_0)\bigr]$&\bmini{3cm}pole\emini\\
&&or&\\
&&$y=\Bigl(\bigl(4/3\bigr)/(x-x_0)\Bigr)^{1/3}$&\bmini{3cm}branch point\emini\\
&&&\\
\hline
\end{tabular}
\end{center}

A normalized ODE, i.e. one that is solved for the highest
derivative, such as
\begin{equation}
y^{(n)}=F\Bigl(y^{(n-1)},\ldots, y', y, z\Bigr),
\end{equation}
gives rise to possible singularities in its solutions
wherever $F$ becomes singular. (Where $F$ is analytic, and
regular initial data are given, standard theorems show
that an analytic solution must exist.) Note that these
singularities may include the points at infinity in $y$ (or its
derivatives) and $z$,
which we denote by $y=\infty$ (or $y'=\infty$ etc), $z=\infty$.
The singularities of $F$, therefore, denote possible
singularities in the solution(s). They may be divided into two
classes: those given by values of $z$ alone and those
involving values of $y$ or its derivatives. The former
are determined \textit{\`a priori} for
all solutions. Therefore, they can only give rise to
\textit{fixed} singularities. To find movable singularities,
we therefore need to investigate the singular values of $F$
that involve
$y$ (or its derivatives). Similar statements can be made
in the case of PDEs.

For example, the Riccati equation
\[y'=-y^2/z =:F(y, z).\]
has a right side $F$ with two singularities given by $z=0$ and
$y=\infty$. The general solution is
\[y(z)={1\over\log(z/z_0)}.\]
It is clear that $z=0$ is a fixed singularity (it stays the
same for all initial conditions) whereas $z_0$ denotes a
movable singularity where $y$ becomes unbounded.

Singularities of nonlinear ODEs need not only occur at points where $y$ is
unbounded. Example 2 of Table 1
indicates possible movable singularities at points where
$y=0$. The solution shows that these are actually movable
branch points.

These considerations show that singular values of the
normalized differential equation lie at the base of
the solutions' singularity structure. Techniques for
investigating singularity structure usually focus on these
singular values.

In the first three subsections below, we describe common
definitions of the \Pa\ property, and the two major techniques
known as the $\alpha$-method and the \Pa\ test for deriving
necessary conditions of the \Pa\ property. In the subsequent
three subsections below, we discuss the need for sufficiency conditions,
the direct method of proving the \Pa\ property, and
convergence-type results for PDEs.

\subsection{The Painlev\'e Property}
The  actual
definition of the
\Pa\ property has been subject to some variation.
There are three  definitions in the literature.
\begin{defn}
An ODE is said to possess
\li{
\item the {\em specialized \Pa\ property} if all movable
singularities of all solutions are poles.
\item the {\em Painlev\'e property} if all solutions are
single-valued around all movable singularities.
\item the {\em generalized \Pa\ property} if the general
solution is single-valued around all movable singularities.}
\end{defn}
\noindent
(The qualifiers \lq\lq specialized\rq\rq\ and \lq\lq generalized\rq\rq\
are not usually used in the literature.)
The first property defined above
clearly implies the others. This property was also the first
one investigated (by ARS) in recent times. It is the property
possessed by the six \Pa\ equations.

The second, more general, definition above is the one used by
\Pa\ in his work on the classification of ODEs. It allows, for
example, movable unbranched essential singularities in any
solution. Of the examples in Table 1, equations 1 and 4 have
the \Pa\ property; equation 1 also has the specialized \Pa\
property. The remaining equations have neither property.

The third property is the most recently proposed variation,
although there is evidence that Chazy assumed it in
investigating ODEs of higher ($\ge 1$) degree or
order ($\ge 2$). The sixth example given in Table 1
satisfies neither of the first two properties above because
the special solution $\bigl(4/3/(x-x_0)\bigr)^{1/3}$ has
movable branch points around which the solution is
multivalued. However, it does satisfy the generalized \Pa\
property because the general solution
$k\tan\bigl[k^3(x-x_0)\bigr]$ is meromorphic.

Most of the known techniques for investigating the \Pa\
property have their origin in the classical work of
Painlev\'e and his colleagues. They classified ODEs of the
form
\begin{equation}
	u''=F(z;u,u'), \label{pclass}
\end{equation}
where F is rational in $u$
and $u'$ and analytic in $z$, according to whether or not
they possess the \Pa\ property.

They discovered that every equation
possessing the Painlev\'e property could either be solved in
terms of known functions (trigonometric functions, elliptic
functions, solutions of linear ODEs, etc.) or transformed to
one of the six equations now called the Painlev\'e equations
($P_I$--$P_{\,V\!I}$). They have standard forms that are
listed below. (They are representatives of
equivalence classes under m\"obius
transformations.) Their general solutions are higher
transcendental functions.
\begin{center}{\bf The Painlev\'e Equations}\end{center}
\begin{eqnarray}
        u'' &=& 6u^2+z,\nn\\
        u'' &=& 2u^3+zu+\A,\nn\\
        u'' &=& \frac 1uu'^2-{1\over z}u'+{1\over z}(\A u^2+\B)+\C u^3
+{\D\over u},\nn\\
        u'' &=& \frac 1{2u}u'^2+\frac 32u^3+4zu^2+2(z^2-\A)u+\frac\B u,\nn\\
        u'' &=& \left\{\frac 1{2u}+\frac 1{u-1}\right\}u'^2
-\frac 1zu'+\frac{(u-1)^2}{z^2}\left(\A+\frac\B u\right)+\frac{\C u}z+
{\D u(u+1)\over u-1},\nn\\
        u'' &=& \half\left\{\frac 1u+\frac 1{u-1}+\frac 1{u-z}\right\}u'^2-
\left\{\frac 1z+\frac 1{u-1}+\frac 1{u-z}\right\}u'\nn\\
        & &  + {u(u-1)(u-z)\over z^2(z-1)^2}
\left\{\A+{\B z\over u^2}+{\C (z-1)\over (u-1)^2}+{\D z(z-1)\over (u-z)^2}
\right\},
\nn
\end{eqnarray}

Two main procedures were used in this work. The first is
known as the $\alpha$-method and the second is now called
\Pa\ analysis. \Pa\ described the $\alpha$-method in the
following way.
\begin{quotation}\textit{
Consid\'erons un \'equation diff\'erentielle dont le
co\"efficient diff\'erentiel est une fonction (holomorphe
pour $\alpha=0$) d'un param\`etre $\alpha$. Si l'equation a
ses points critiques fixes pour $\alpha$ quelconque (mais
$\not=0$), il en est de m\^eme, a fortiori pour $\alpha=0$,
et le d\'eveloppement de l'int\'egrale $y(x)$, suivant les
puissances de $\alpha$, a comme co\"efficients des fonctions
de $x$ \`a points critiques fixes.}
\end{quotation}
(This extract is taken from footnote 3 on p.11 of
\cite{painleve02}. In \Pa's terminology, a critical point of
a solution is a point around which it is multivalued.) In
other words, suppose a parameter $\alpha$ can be introduced
into an ODE in such a way that it is analytic for
$\alpha=0$. Then if the ODE has the \Pa\ property for
$\alpha\not=0$, it must also have this property for
$\alpha=0$. We illustrate this method for the classification
problem for first-order ODEs below.

The second procedure, called \Pa\ analysis, is a method of
examining the solution through formal expansions in
neighbourhoods of singularities of the ODE. In particular,
the procedure focusses on formal series expansions of the
solution(s) in neighbourhoods of generic (arbitrary) points
(not equal to fixed singularities). The series expansion is
based on Fr\"obenius analysis and usually takes the form
given by Eqn\rf{pseries}.

As mentioned in the Introduction, this procedure was
extended to PDEs by WTC (Weiss, Tabor, and Carnevale)
\cite{weisstc}.  For PDEs, the above definitions
of the \Pa\ property continue to hold under the interpretation
that a
\textit{movable singularity} means
\textit{noncharacteristic analytic movable
singularity manifold}.

A noncharacteristic manifold for a given PDE is a surface on which we can
freely specify Cauchy data.  The linear wave equation,
\begin{equation}
        u_{tt}-u_{xx}=0,        \label{wave}
\end{equation}
has the general solution
$$      u(x,t)=f(t-x)+g(t+x),   $$
where $f$ and $g$ are arbitrary.  By a suitable choice of $f$ and $g$ we can
construct a solution $u$ with any type of singularity on the curves
$t-x=k_1$, $t+x=k_2$, for arbitrary constants $k_1$, and $k_2$.
These lines are characteristic manifolds for
equation~\rf{wave}.  This example illustrates why the Painlev\'e property
says nothing about the singular behaviour of solutions on characteristic
singularity manifolds.

The WTC procedure is to expand the solutions $u(x,t)$ of a PDE as
\beqn
u(x,t)=\sum_{n=0}^{\infty}u_n(x,t) \Phi^{n+\rho},
\label{pdeseries}
\eeqn
near a
noncharacteristic analytic movable
singularity manifold given by $\Phi=0$.
(This extends in the obvious way for functions of more
than two variables.) The actual expansion can be simplified
by using information specific to the PDE about its characteristic
directions. For example, for the KdV equation, noncharacteristic means
that $\Phi_x\not=0$. Hence by using the implicit function theorem
near the singularity manifold, we can write
\[\Phi(x, t)= x-\xi(t),\]
where $\xi(t)$ is an arbitrary function.
This is explored further in subsection 2.3.2 below.

In some cases, the form of the series Eqn\rf{pseries} (or (\rf{pdeseries})
needs modification. A simple example of this is the ODE
\beqn
u'''=2(u')^3+1.\label{jacoprime}
\eeqn
Here $v=u'$ is a Jacobian elliptic function with simple
poles of residue $\pm 1$. (See \cite{abramowitzs}.) Hence a
series expansion of $u(z)$ around such a singularity $z_0$,
say, must start with $\pm\log(z-z_0)$. The remainder of the
series is a power series expansion in powers of $z-z_0$.
In such cases, the \Pa\ property holds for the new variable
$v$.

\subsection{The $\alpha$-Method}
In this section, we illustrate the $\A$-method by using it
to find all ODEs of the form
\begin{equation}
	u'={P(z,u)\over Q(z,u)}		\label{firstpain}
\end{equation}
possessing the Painlev\'e property, where $P$ and $Q$
are analytic in $z$ and polynomial in
$u$ (with no common factors).  The first step of the
$\A$-method is to introduce a small parameter $\A$
through a transformation of variables in such a way
that the resulting ODE is analytic in $\A$. However,
the transformation must be suitably chosen so that the
limit $\A\to0$ allows us to focus on a movable
singularity. This is crucial for deducing necessary
conditions for the \Pa\ property.

We accomplish this by using dominant balances of the ODE
near such a singularity. (See \cite{kruskal63,bendero} for a
definition and discussion of the method of dominant
balances.)

If $Q$ has a zero of multiplicity $m$ at
$u=a(z)$ then, after performing the transformation
$u(z)\mapsto u(z)+a(z)$, equation~\rf{firstpain} has the form
\begin{equation}
	u^mu'=f(z,u),	\label{firsttrans}
\end{equation}
where $f$ is analytic in $u$ at $u=0$ and $f(z,0)\not\equiv0$ (since $P$ and
$Q$
have no common factors).  Choose $z_0$ so that $\kappa:=f(z_0,0)\ne0$ and
define the transformation
$$	u(z)=\alpha U(Z),\quad z=z_0+\alpha^n Z, $$
where $n$ is yet to be determined and $\alpha$ is a small (but nonzero)
parameter. Note that this is designed to focus on solutions that become
close to the singular value $u=0$ of the equation somewhere in the $z$-plane.

Equation~\rf{firsttrans} then becomes
$$
\alpha^{m+1-n}U^m{dU\over dZ}=
		f(z_0+\alpha^nZ,\alpha U)=\kappa+O(\alpha).
$$
This equation has a dominant balance when $n=m+1$.  In this case the limit
as $\alpha\to0$ gives
$$
	U^m{dU\over dZ}=\kappa
$$
which has the exact solution
$$
U(Z)=\left\{ (m+1)\kappa Z+C\right\}^{1\over m+1},
$$
where $C$ is a constant of integration.
This solution has a movable branch point at $Z=-C/\bigl(\kappa(m+1)\bigr)$
for all $m>0$.
Therefore, Eqn~\rf{firstpain} cannot possess the Painlev\'e
property unless $m=0$, i.e. Q must be independent of
$u$.  That is, to possess the \Pa\ property,
Eqn~\rf{firstpain} must necessarily be of the form
\begin{equation}
	u'=a_0(z)+a_1(z)u+a_2(z)u^2+\cdots+a_N(z)u^N,
\label{firstred}
\end{equation}
for some nonnegative integer $N$.

The standard theorems of
existence and uniqueness fail for this equation wherever $u$
becomes unbounded. To investigate what happens in this case,
we transform to $v:=1/u$. (Note that the \Pa\ property is
invariant under such a transformation.) Eqn~\rf{firstred} then
becomes
$$
v'+a_0(z)v^2+a_1(z)v+a_2(z)+a_3(z)v^{-1}+\cdots
+a_N(z)v^{2-N}=0.
$$
But this equation is of the form \rf{firstpain} and so it can
only possess the Painlev\'e property if $N=2$.

In summary, for Eqn~\rf{firstpain} to possess the
Painlev\'e property, it must necessarily be a Riccati
equation:
\begin{equation}
	u'=a_0(z)+a_1(z)u+a_2(z)u^2. 	\label{riccati}
\end{equation}
To show that this is also sufficient, consider the
transformation
$$
u=-\frac1{a_2(z)}{w'\over w}
$$
which linearizes Eqn~\rf{riccati}
$$ a_2w''-(a_2'+a_1a_2)w'+a_0a_2^2w=0. $$
By Fuchs theorem \cite{bendero},
the singularities of any solution
$w$ can only occur at the singularities of
$(a_2'+a_1a_2)/a_2$ or
$a_0a_2$. These are fixed singularities. Hence the only
movable singularities of $u$ occur at the zeroes of $w$.
Since $w$ is analytic at its zeroes, it follows that $u$ is
meromorphic around such points. That is, Eqn~\rf{riccati}
has the \Pa\ property.

\subsection{The Painlev\'e Test}
Here we illustrate the widely used formal tests for the
\Pa\ property for ODEs and PDEs by using examples.

\subsubsection{ODEs}
Consider a class of ODEs given by
\begin{equation}
	u''=6u^n+f(z),	\label{secondpain}
\end{equation}
where $f$ is (locally) analytic and $n\ge 1$ is an integer
(the cases
$n=0$ or $1$ correspond to linear equations).

Standard theorems that yield analytic solutions
fail for this equation wherever the right side becomes
singular, i.e. where either $f(z)$ or $u$ becomes unbounded.
We concentrate on the second possibility to find movable
singularities. This means that the hypothesized expansion
Eqn\rf{pseries} must start with a term that blows up at
$z_0$. To find this term, substitute
$$	u(z) \sim c_0(z-z_0)^p,\qquad z\to z_0, $$
where $\Re(p)<0$, $c_0\ne 0$, into
Eqn~\rf{secondpain}. This gives the dominant equation
\begin{equation}
	c_0p(p-1)(z-z_0)^{p-2} + O\left( (z-z_0)^{p-1}\right) =
		6c_0^2(z-z_0)^{np}+O\left((z-z_0)^{np+1}\right).
			\label{secondexp}
\end{equation}
The largest terms here must balance each other (otherwise
there is no such solution). Since
$c_0\ne0$ and $p\ne0$ or $1$, we get
$$
p-2=np,\quad\Rightarrow\quad p={-2\over n-1}.
$$
If $p$ is not an integer, then $u$ is branched at $z_0$.
Hence, the only
$n>1$ for which equation~\rf{secondpain} can possess the Painlev\'e property
are $n=2$ and $n=3$.

We will only consider the case $n=2$
here for conciseness.  The case $n=3$
is similar. (The reader is urged to retrace the following
steps for the case $n=3$.)

If $n=2$ then $p=-2$ (which is
consistent with our assumption that $\Re(p)
<0$). Then equation~\rf{secondexp} becomes
$$
6c_0(z-z_0)^{-4} = 6c_0^2+O\left( (z-z_0)^{-3}\right),
$$
which gives $c_0=1$.  Hence the hypothesized series expansion for $u$ has
the form
\begin{equation}
	u(z)=\sum_{n=0}^\infty c_n(z-z_0)^{n-2}.\label{secondseries}
\end{equation}

The function $f(z)$ can also be expanded in a
power series in $z-z_0$ because by assumption it is analytic.
Doing so and substituting expansion
\rf{secondseries} into eqn\rf{secondpain} gives
\begin{eqnarray}
& & \sum_{n=0}^\infty (n-2)(n-3)c_n(z-z_0)^{n-4} \nn\\
&=& 6\sum_{i,j=0}^\infty c_ic_j(z-z_0)^{i+j-4}+
\sum_{m=0}^\infty {1\over m !} f^{(m)}(z_0)(z-z_0)^m \nn\\
&=& 6(z-z_0)^{-4}+12c_1(z-z_0)^{-3}\nn\\
& & +6(c_1^2+2c_2)(z-z_0)^{-2} +12(c_3+c_1c_2)(z-z_0)^{-1}\nn\\
& & + \sum_{n=4}^\infty\left\{ 6\sum_{m=0}^n c_mc_{n-m}+
{1\over (n-4)!} f^{(n-4)}(z_0)\right\}(z-z_0)^{n-4}.\nn
\end{eqnarray}
Equating coefficients of like powers of $(z-z_0)$
we get $c_1=0$, $c_2=0$,
$c_3=0$, and
$$ (n-2)(n-3)c_n=6\sum_{m=0}^n c_mc_{n-m}+
{1\over (n-4)!} f^{(n-4)}(z_0),
\quad (n\ge 4). $$
Note that $c_n$ appears on both sides of this equation. Solving for $c_n$,
we find
\begin{equation}
 (n+1)(n-6)c_n =6\sum_{m=1}^{n-1}c_mc_{n-m}+
{1\over (n-4)!}f^{(n-4)}(z_0).
			\label{recur}
\end{equation}
For each
$n\ne 6$, this relation defines $c_n$ in terms of
$\{c_m\}_{0\le m<n}$.  However, for $n=6$, the coefficient of
$c_n$ vanishes and equation~\rf{recur} fails to define $c_6$.
If the right side also vanishes, $c_6$
is arbitrary. However, if the right side does not vanish
there is a contradiction which implies that the series \rf{secondseries}
cannot be a formal solution of eqn\rf{secondpain}.

In that second case, the expansion can be modified to
yield a formal solution by inserting appropriate logarithmic terms starting
at the index $n=6$. (This is also the case for Fr\"obenius expansions
when the indicial exponents differ by an integer -- see \cite{bendero}.
See \cite{kichenassamys} also for a rigorous study of equations admitting
such algebraico-logarithmic expansions in several variables.)
In such a case, logarithmic terms appear infinitely often in the
expansion and cannot be transformed away (as in Eqn\rf{jacoprime} above).
They therefore indicate multivaluedness around movable singularities.

That is, Eqn\rf{secondpain} fails the Painlev\'e test
unless the right side of Eqn\rf{recur} vanishes at $n=6$.
This condition reduces to
$$
f''(z_0)=0.
$$
However, since $z_0$ is arbitrary, this implies $f''\equiv 0$. That is,
$f(z)=az+b$ for some constants $a$, $b$.
If $a=0$, this equation can be solved in terms of (Weierstrass)
elliptic functions.  Otherwise, translating $z$ and rescaling $u$ and $z$
gives
\begin{equation}
	u''=6u^2+z,	\label{pi}
\end{equation}
which is the first Painlev\'e equation.

The index of the free coefficient, $c_6$, in the above expansion is
called a {\em resonance}. The expansion contains two
arbitrary
constants, $c_6$ and $z_0$, which indicates that it captures
the generic singular behaviour of a solution (because the
equation is second order).

There is a standard method for
finding the location of resonances which avoids
calculation of all previous coefficients. We illustrate
this method here for $P_I$.
After determining the leading order behaviour, substitute the perturbation
$$
u \sim (z-z_0)^{-2}+\cdots+\beta (z-z_0)^{r-2}
$$
where $r>0$
into equation~\rf{pi}. Here $\beta$ plays the
role of the arbitrary coefficient. To find a resonance $r$,
we collect terms in the equation that are linear in $\beta$ and
demand that the coefficient of $\beta$ vanishes. This is equivalent
to demanding that $\beta$ be free.
The resulting equation
$$
(r+1)(r-6)=0,
$$
is called the resonance equation and is precisely the coefficient
of $c_r$ on the left side of equation \rf{recur}.

The positive root $r=6$ is precisely the
resonance we found earlier. The negative root $r=-1$, often
called the universal resonance, corresponds to the translation
freedom in $z_0$. (Consider $z_0\mapsto z_0+\epsilon$.
Taking $|\epsilon|<|z-z_0|$, and expanding in $\epsilon$
shows that $r=-1$ does correspond to an arbitrary
perturbation.)

Note, however, that $r=-1$ is not always a resonance.
For example, consider an expansion that starts with a nonzero
constant
term, such as
\[ 1+a_1(z-z_0)+\ldots .\]
Perturbation of $z_0$ does not add a term corresponding
to a simple pole to this expansion.

If any resonance is not an integer, then the equation fails
the Painlev\'e test.  The role played by other
negative integer resonances is not fully understood. We
explore this issue further through irregular singular point
theory in Section 3.

For each resonance, the \textit{resonance condition} needs
to be verified, i.e. that the equation at that index is
consistent. These give rise to necessary conditions for the \Pa\
property. If all nonnegative resonance conditions are satisfied
and all formal solutions around all generic arbitrary
points $z_0$ are meromorphic, the equation is said to pass the
\Pa\ test.

This procedure needs to be carried out for evey possible singularity
of the normalized equation.
For example, the sixth
Painlev\'e equation,
$P_{\,V\!I}$, has four singular values in $u$, i.e. $u=0$,
$1$, $z$ and $\infty$.  The expansion procedure outlined
above needs to be carried out around arbitrary points where
$u$ approaches each such singular value.
(Table~2 in Section 4 lists all singular values of
the Painlev\'e equations.)

\subsubsection{PDEs}
In this subsection, we illustrate the WTC series expansion technique
with an example. Consider the
variable coefficient KdV equation,
\begin{equation}
u_t+f(t)uu_x+g(t)u_{xxx}=0.\label{varkdv}
\end{equation}
Let $\phi(x,t)$ be an arbitrary holomorphic function such that
$S:=\{(x,t)\,:\,\phi(x,t)\}=0$ is
noncharcteristic.
The fact that $S$ is noncharacteristic
for equation~\rf{varkdv} means that
\begin{equation}
\phi_x\ne0      \label{implicit}
\end{equation}
on $S$. By the implicit function theorem,
we have locally
$\phi(x,t) = x-\xi(t)$
for some arbitrary function $\xi(t)$.

We begin by substituting an expansion of the form
\begin{equation}
u(x,t)=\sum_{n=0}^{\infty}u_n(t)\phi^{n+\alpha}
\end{equation}
into equation~\rf{varkdv}.  The leading order terms give
$\alpha=-2$.  Equating coefficients of powers of $\phi$ gives
\begin{eqnarray}
n=0&:&u_0 = -12 g/f,\\
n=1&:&u_1 =0,\\
n=2&:&\ u_2 = \xi'/f\\
n=3&:&\ u_3 = u_0'/(fu_0),\\
n\ge 4&:&(n+1)(n-4)(n-6)g u_n \\
& &\qquad = -f\sum_{k=0}^{n-4}(k+1)u_{n-k-3}u_{k+3}\\
 & &\qquad\qquad +(n-4)\xi'u_{n-2}-u'_{n-3}
\end{eqnarray}
Arbitrary coefficients can enter at $n=4$, $6$ if
the recursion relation is consistent.
Consistency at $n=6$ is equivalent to
$$
\left({u_0'\over fu_0}\right)^2 +{1\over f}\left({u_0'\over fu_0}\right)_t=0.
$$
This implies that
\[g(t)=f(t)\left\{a_0\int^t f(s)ds+b_0\right\},
\]
where $a_0$ and $b_0$ are arbitrary constants.
In this case, Eqn\rf{varkdv}
can be transformed exactly to the KdV equation
(see Grimshaw~\cite{grimshaw}).

In particular, for the usual form of the KdV ($f(t)=6$, $g(t)=1$) we have
the formal series expansion
\begin{eqnarray}
u(x,t) &=& {-2\over \phi^2}+{\xi'(t)\over 6}+u_4(t)\phi^2\nn\\
& &
\ -{\xi''(t)}\phi^3+u_6(t)\phi^4+O\bigl(\phi^5
\bigr).        \label{kdvexpansion}
\end{eqnarray}
Questions of convergence and well-posedness, i.e. continuity of the solution
as the arbitrary functions ($\xi(t)$,
$u_4(t)$, $u_6(t)$) vary, are discussed in Section 2.6 below.

\subsection{Necessary versus sufficient conditions for the \Pa\ property}
The methods we described above can only yield necessary conditions for
the \Pa\ property.
Here we illustrate this
point with an example (due to \Pa )
which does not possess the \Pa\ property, but
for which the \Pa\ test indicates only meromorphic
solutions.

Consider the ODE
\beqn (1+u^2)u''+(1-2u)u'^2=0 \label{dangerous}\eeqn
(see Ince~\cite{ince}).  The singularities of this equation are $u=\pm i$,
$u=\infty$ and $u'=\infty$.
Series expansions can be developed for solutions exhibiting
each of the above singular behaviours and the equation passes the Painlev\'e
test.  This equation,
however, has the general solution
$$u(z)=\tan \{\log [k(z-z_0)]\},$$
where $k$ and $z_0$ are constants.  For $k\ne0$, $u$ has poles at
$$	z=z_0+k^{-1}\exp\{-(n+1/2)\pi\} $$
for every integer $n$.
These poles accumulate at the movable point $z_0$, giving rise to a movable
branched nonisolated essential singularity there.  This example clearly
illustrates the fact that passing the Painlev\'e test is not a guarantee that
the equation actually possesses the Painlev\'e property.

This danger arises also for PDEs. The PDE
$$
w_t=(1+w^2)w_{xx}+(1-2w)w_x,
$$
under the assumption
$$
w(x,t)=:u(x)
$$
reduces to the ODE above.

To be certain that a given differential equation possesses the \Pa\ property,
we must either solve it explicitly or implicitly (possibly through
transformations to other equations known to have the property), or develop
tests for sufficiency. Most results in the literature rely on the
former approach. In the next section, we develop a method for testing
sufficiency.

\subsection{A Direct Proof of the Painlev\'e Property for ODEs}
In this section we outline a direct proof given in (Joshi and
Kruskal~\cite{joshik92,joshik94}) that the Painlev\'e equations indeed
possess the Painlev\'e property. The proof is based
on the well known Picard iteration method (used to prove
the standard theorems of existence and analyticity of solutions
near regular points) modified to apply near singular
points of the \Pa\ equations. A recommended simple example
for understanding the method of proof is
\[u''=6u^2 + 1\]
which is solved by Weierstrass elliptic functions.

Consider the initial value problem for each of the six \Pa\ equations
with bounded data for $u$ and $u'$ given at an ordinary point $z_1\in\Complex$.
(The point $z_1$ cannot equal $0$ for $P_{III}-P_{VI}$, $1$ for $P_V$ or
$P_{VI}$, or $z_1$ for $P_{VI}$ --- see Table 2.)
\begin{center}
\begin{tabular}{|c|c|c|}
\hline
\multicolumn{3}{|c|}{\bf Table 2: Fixed and Movable Singularities}\\
\multicolumn{3}{|c|} {\bf of the
Painlev\'e Equations}\\
\hline
Equation & Fixed Singularities & Movable Singularities\\
        &   ($z$-value) &       ($u$-value)     \\
\hline\hline
$ P_I$ &
$ \{\infty\} $ &      $ \{\infty\} $ \\
\hline
$P_{II}$        &$      \{\infty\} $ & $ \{\infty\} $\\
\hline
$P_{III}$       &        $\{0,\infty\} $ & $ \{0,\infty\} $ \\
\hline
$P_{IV}$ & $ \{\infty\}$ & $ \{0,\infty\} $ \\
\hline
$P_{\,V}$ & $\{0,\infty\} $ & $ \{0,1,\infty\} $ \\
\hline
$P_{\,V\! I}$ & $\{0,1,\infty\} $ & $ \{0,1,z,\infty\} $ \\
\hline
\end{tabular}
\end{center}

Standard theorems yield a (unique) solution $U$ in any region in which
the Lipschitz condition holds.  However, they fail where the right
side becomes unbounded, i.e. at its singular values. (See e.g.
\cite{coddingtonl}.) Since our purpose is to study the behaviour
of the solution near its movable singularities, and these lie
in the finite plane, we confine our attention to an arbitrarily
large but bounded disk $|z|<B$ (where $B$ is real and say $>1$). For
$P_{III}$, $P_V$, and $P_{VI}$ this must be punctured at the finite
fixed singularities. Henceforth we concentrate on $P_I$ for simplicity.

The ball $|z|<B$ contains two types of regions. Around each movable
singularity, we select a neighbourhood where the largest terms in
the equation are sufficiently dominant over the other terms.
We refer to these as \textit{special regions}. Outside these special
regions, the terms remain bounded. Therefore, the ball resembles
a piece of Swiss cheese, the holes (which may not be circular
in general but in this case turn out to be nearly circular)
being the special regions where movable singularities reside and the solid
cheese being free of any singularity.

Starting at
$z_1$ in (the cheese-like region of) the ball,
we continue the solution $U$ along a ray until we encounter a point
$z_2$ on the edge of a special region. Inside the region, we convert
$P_I$ to an integral equation by operating successively on the equation
as though only the dominant terms were present.

The dominant terms of
\beqn u''=6u^2+z,\label{piagain}\eeqn
are $u''$ and $6u^2$. Integrating these dominant terms after multiplying
by their integrating factor $u'$, we get
\begin{equation}
{u'^2\over 2} = 2u^3+zu-\int_{z_2}^{z}u\,dz+\bar k, \label{piint1}
\end{equation}
with
$$ \bar k :=\int_{z_1}^{z_2}u\,dz+k,	$$
where the constant $k$ (kept fixed below)
is determined explicitly by the initial conditions.

Since $u$ is large, $u'$ does not vanish (according to Eqn\rf{piint1})
and, therefore,
there is a path of steepest ascent starting at $z_2$. We will use
this idea to find a first point in the special region where $u$ becomes
infinite.

Let $d$ be an upper bound on
the length of the path of integration from
$z_2$ to $z$ and assume $A>0$ is given such that
\[A^2>4B, A^2>4\pi B, A^2>4d, A^3>4|k|.\]
Assume that $|u|\ge A$ at $z_2$. Then Eqn\rf{piint1} gives
$u'(z_2)\ne=0$. Taking the path of integration to be the path
of steepest ascent, we can show that
\[|u'|>\sqrt{2}|u|^{3/2},\]
and that the distance to a point where $|u|$ becomes infinite is
\[ d<\sqrt{2}A^{-1/2}.\]
(See page 193 of \cite{joshik94} for details.)
So there is a first singularity encountered on this path which
we will call $z_0$.

Now integrating the dominant terms once more (by dividing by
$2u^3$, taking the square root and integrating
from $z_0$) we get
\begin{equation}
u=\left(\int_{z_0}^{z}\left\{
	1+{1\over 2u^3}\left(\bar k+zu-\int_{z_2}^z u\,dz\right)
	\right\}^{1/2}dz\right)^{-2}. \label{uint}
\end{equation}

Substituting a function of the form
$$u(x,t)={1\over (z-z_0)^2} + f(z)$$
where $f(z)$ is analytic at $z_0$
into the right side of equation~\rf{uint} returns a function of the same form.
Notice that, therefore, no logarithmic terms can arise.

It is worth noting that the iteration of the integral equation\rf{uint}
gives the same expansion that we would have obtained by the \Pa\ test.
In particular, it generates the appropriate formal solution without
any assumptions of its form and it point outs precisely
where logarithmic terms may
arise without additional investigations. (For example, try iteration
of the integral
equation with the term $z$ on the right side of equation\rf{piagain}
replaced by $z^2$,
i.e. with $zu-\int_{z_2}^z u\,dz$ replaced by
$z^2u-2\int_{z_2}^z zu\, dz$ in Eqn\rf{uint}.)

The remainder of the proof is a demonstration that the integral equation
\rf{uint} has a unique solution meromorphic in a disk centred at
$z_0$, that its radius is lower-bounded by a number that
is independent of $z_0$, and that the solution is the same as
the continued solution $U$. The uniformity of the lower bound is crucial
for the proof. Uniformity excludes the possibility
that the movable poles may accumulate to form movable essential
singularities as in example\rf{dangerous}.

Since the analytic continuation of $U$
is accomplished along the union of segments of rays and circular
arcs (skirting around the
boundaries of successive special regions encountered on such
rays) and these together with the special regions cover the whole ball
$|z|<B$, we get a proof that the first \Pa\ transcendent is meromorphic
throughout the ball.

\subsection{Rigorous Results for PDEs}
Sufficient results for the \Pa\ property of PDEs
have been harder to achieve than ODEs. This is surprising
because such results are lacking even for the most well
known integrable PDEs. In this section,
we describe some partial results towards this direction
for the KdV equation.

\begin{defn} The {\em WTC data} for the KdV equation
is the set $\{\xi(t)$, $u_4(t)$, $u_6(t)\}$
of arbitrary functions describing this Painlev\'e expansion.
\end{defn}
The following theorem proves that the series \rf{kdvexpansion} converges
for analytic WTC data.
\begin{theorem}
 (Joshi and Petersen~\cite{joship1,joship2})
Given an analytic manifold
$S:=\{(x,t): x = \xi (t) \}$,
with
$\xi(0) = 0$, and two arbitrary analytic functions
$$
\lim_{x\to\xi(t)}\Bigl(\frac{\partial\ }{\partial
x}\Bigr)^4[w(x,t)(x-\xi(t))^2], \quad
\lim_{x\to\xi(t)}\Bigl(\frac{\partial\ }{\partial
x}\Bigr)^6[w(x,t)(x-\xi(t))^2]
$$
there exists in a neighbourhood of $(0,0)$ a
meromorphic solution of the KdV equation~\rf{kdv}
of the form
$$
w(x, t) = \frac{-12}{\bigl(x-\xi(t)\bigr)^2} + h(x, t)
$$
where $h$ is holomorphic.
\end{theorem}
\noindent The next theorem provides us with a useful lower bound on the
radius of convergence of this series.
\begin{theorem} (Joshi and Srinivasan~\cite{joshis})
Given  WTC data $\xi(t)$, $u_4(t)$, $u_6(t)$ analytic in the ball
$B_{2\rho+\epsilon}(0)= \{t\in {\bf C} :$ $ |t| < 2\rho +\epsilon\}$,
let
$$
M = \sup_{|t| = 2\rho}\{1,  |\xi(t)|,
|u_4(t)|, |u_6(t)|\}.
$$
The radius of convergence $R_{\rho} = R$ of the power series (1.5)
satisfies
$$
R \geq \frac{\min\{1,\rho\}}{10M}.
$$
\end{theorem}
This lower bound is used in \cite{joshis} to prove the well-posedness
of the WTC Cauchy problem.  That is, the locally meromorphic function function
described by the convergent series~\rf{kdvexpansion} has continuous dependence
on the WTC data in the sup norm.

To date there is no proof that the
Korteweg-de~Vries equation possesses
the Painlev\'e property. The main problem lies in a lack of
methods for obtaining the global
analytic description of a locally defined solution in the
space of several complex variables. However, some partial results
have been obtained.

The usual initial value problem for the KdV equation is given on
the characteristic manifold $t=0$. Well known symmetry reductions of
the KdV equation (e.g. to a \Pa\ equation) suggest that a generic
solution must possess an infinite number of poles. WTC-type analysis
shows that these can occur on noncharacteristic manifolds which
intersect $t=0$ transversely. These results suggest that only
very special solutions can be entire (i.e. have no singularities)
on $t=0$.

Joshi and Petersen \cite{joship96} showed that if the initial value
$$
u(x,0)=u_0(x):= \sum^{\infty}_{n=0} a_n x^n,
$$
is entire in $x$ and, moreover, the coefficients $a_n$ are real
and nonnegative then there exists no solution holomorphic in any
neighbourhood of the origin in $\Complex^2$ unless
\[u_0(x) = a_0+a_1x.\]
This result can be extended to the case of more general $a_n$
under a condition on the growth of the function $u_0(x)$ as $x\to\infty$.

\section{Nonlinear-Irregular-Singular Point Analysis}
The \Pa\ expansions cannot describe all possible singular behaviours of
solutions of differential equations. In this section, we describe some
extensions based on irregular-singular-point theory for linear equations.

The \Pa\ expansions at their simplest are Laurent series with
a leading
term, and can, therefore, fail to describe solutions
that possess movable isolated essential singularities.
Consider the ODE
$$
3u'u'''=5(u'')^2-(u')^2{u''\over u}-{(u')^4\over u^2},
$$
which has the general solution
$$	u(z)=\A\exp\left\{\B (z-z_0)^{-1/2}\right\}.	$$
Clearly $u$ has a branched movable essential singularity.
As suggested in \cite{kruskal}, the Painlev\'e test can be extended to capture
this behaviour by considering solutions that become exponentially large
near $z_0$.  To do this we expand
$$
u(z)=a_{-1}(z){\rm e}^S(z)+a_0(z)+a_1(z){\rm e}^{-S(z)}
 + a_2(z){\rm e}^{-2S(z)}+a_3(z){\rm e}^{-3S(z)}+\ldots,
$$
where $S$ and the $a_n$ are generalized power series and $S$ grows faster than
any logarithm as $z$ approaches $z_0$.

In other words, generalized expansions (those involving logarithms,
powers, exponentials and their compositions) are necessary if we are
to describe all possible singularities. These are asymptotic expansions
which may fail to
converge. They are in fact asymptotic expansions. We show in this section
that they can nevertheless
yield analytic information about solutions. We illustrate
this with two main examples.
The first is the Chazy equation and the second a fourth-order
equation studied by Bureau.

\subsection{The Chazy Equation}
In this subsection, we
examine the Chazy equation
\begin{equation}
y'''=2yy''-3(y')^2.	\label{chazy}
\end{equation}
This equation is exactly solvable through the transformation~
\cite{chazy1,chazy2}
$$
z(t):={u_2(t)\over u_1(t)},\quad
   y(z(t))={6\over u_1(t)}{du_1(t)\over dt}=6{d\ \over dt}\log u_1(t),
$$
where $u_1$ and $u_2$ are two independent solutions of the Hypergeometric
equation
$$
t(t-1){d^2u\over dt^2}+\left( {1\over 2}
- {7\over 6}t\right) {du\over dt} -{u\over 144}=0.
$$

Following the work of Halphen~\cite{halphen}, Chazy noted that the function
$z(t)$ maps the upper half $t$-plane punctured at $0$, $1$, and $\infty$
to the interior of a circular triangle with angles $\pi/2$, $\pi/3$, and
$0$ in the $z$-plane (see, for example, Nehari~\cite{nehari}).  The analytic
continuation of the solutions $u_1$ and $u_2$ through one of the intervals
$(0,1)$, $(1,\infty)$, or $(-\infty,0)$ corresponds to an inversion of
the image triangle across one of its sides to a complementary triangle.
Continuing this process indefinitely leads to a tessellation
of either the
interior or the exterior of a circle on the Riemann sphere
This circle is a
{\em natural barrier} in the sense that the solution can be analytically
extended up to but not through it.

We will see below that any solution of equation~\rf{chazy} is
single-valued everywhere it is defined.  The general solution, however,
possesses a movable natural barrier, i.e. a closed curve on the
Riemann sphere whose location depends on initial conditions
and through which the
solution cannot be analytically
continued.

Leading order analysis of equation~\rf{chazy} shows that near a pole,
$$	y\sim -6(z-z_0)^{-1},\quad\mbox{or}\quad
	y\sim A(z-z_0)^{-2},	$$
where $A$ is an arbitrary (but nonzero) constant.
On calculating successive terms in this generalized series expansion we
find only the exact solution
\begin{equation}
	y(z)={A\over (z-z_0)^2}-{6\over z-z_0}.	\label{exact}
\end{equation}
This solution has
only two arbitrary constants, $A$ and $z_0$, and clearly
cannot describe all possible solutions of equation~\rf{chazy}
which is third order.
That is, solutions of the form \rf{exact} fail to capture
the generic behaviour of the full space of solutions.

The absent degree of freedom may lie in a
perturbation of this solution.
Applying the usual procedure for locating resonances
i.e. substituting the expression
$$	y(z)\sim {-6\over (z-z_0)^2}+\cdots+\beta (z-z_0)^{r-2}	$$
into equation~\rf{chazy} and demanding that $\beta$ be free,
we find
$$(r+1)(r+2)(r+3)=0.$$
I.e. we must have $r=-1$, $r=-2$ or $r=-3$.  The case $r=-1$
corresponds to the fact that $A$ is arbitrary in \rf{exact}.  The case $r=-2$
corresponds to the freedom in $z_0$.
The case $r=-3$, however, indicates something more.

Since the usual Fr\"obenius-type series
fails to describe the general solution near a singular
point, we turn to be a nonlinear analogue of irregular
singular point theory.  Arguing from analogy with the linear theory (see, for
example, Bender and Orszag~\cite{bendero}) we seek a solution of the form
\begin{equation}
	y(z)={A\over (z-z_0)^2}-{6\over (z-z_0)} + \exp S(z),	\label{irreg}
\end{equation}
where $\exp S(z)$ is regarded as small in a region near $z_0$ (generically,
$z_0$ will be on the boundary of this region).

Since the Chazy equation is autonomous we can, without loss of generality, take
$z_0=0$. For simplicity we also take $A=1/2$.
Substituting the expansion~\rf{irreg}
into equation~\rf{chazy} gives
\begin{eqnarray}
& &S'''+3S'S''+S'^3\nn\\
& &\quad =
	\left({1\over z^2}-{12\over z}\right)\left( S''+S'^2\right)
        +6\left({1\over z^3}-{6\over z^2}\right)S'\nn\\
& &\quad\  +6\left({1\over z^4}-{4\over z^3}\right)+2\left( S''+S'^2
\right){\rm e}^S
-3S'^2{\rm e}^S. \label{Seqn}
\end{eqnarray}
To ensure that $\exp(S)$ is exponential rather than algebraic, we must
assume that $S''\ll S'^2$, $S'''\ll S'^3$.
Using these assumptions along with $\exp S\ll1$, $S'\gg1$, and $z\ll1$,
equation~\rf{Seqn} gives
$$	S'\sim {1\over z^2}-{2\over z}.$$
Integration yields
$$ y(z) \sim {1\over 2z^2}-{6\over z} +{k\over z^2}{\rm e}^{-1/z}, $$
where $k$ is an arbitrary constant; $k$ represents the third
degree of freedom that was missing from the Laurent
series\rf{exact}. Extending to higher orders in $\exp S(z)$, we obtain
the double series
\begin{eqnarray}
	y(z) &=& {1\over 2z^2}-{6\over z}\nn\\
& & +{k\over z^2}{\rm e}^{-1/z}\left(1+O(z)\right) +{k^2\over 8z^2}
{\rm e}^{-2/z}\left(1+O(z)\right)\label{chazyseries}\\
& & \quad +O\left({{\rm e}^{-3/z}\over z^2}\right).\nn
\end{eqnarray}
It can be shown that this series is convergent in a half-plane, given
here by $\Re(1/z)>0$.

This asymptotic series is valid wherever
\begin{equation}
		|k\exp(-1/z)|\ll1.\label{validity}
\end{equation}
Suppose $k$ is small. Put
$$	z=-\xi+\eta,	$$
where $\xi>0$ to see whether the half-plane of validity can be
extended.  Then the condition~\rf{validity} becomes
$${\xi\over \xi^2+\eta^2}\ll\log\left({1\over |k|}\right). $$
By completing squares (after multiplying out the denominator)
this can be rewritten as
$$
(\xi-\D)^2+\eta^2 \gg\D^2,
$$
where
$$	\delta:=-{1\over 2\log |k|} > 0. $$
So asymptotically the region of validity of the series \rf{chazyseries} lies
outside a circle of radius $\D$ centered at $-\D$.  This is the circular
barrier present in the general solution of the Chazy equation.
In summary, the exponential (or WKB-type) approach has led to a three
parameter solution. Morover, this description is valid in a region
bounded by a circular curve where it diverges.

\subsection{The Bureau Equation}
Bureau partially extended the classification work of \Pa\ and colleagues to
fourth-order equations. However, there were cases
whose \Pa\ property could not be determined within the class of
techniques developed by \Pa's school. One of these was
\beqn
u''''=3u''u-4u'^2, \label{bureau7}
\eeqn
which we will call the Bureau equation. In this subsection, we show that the
general solution is multivalued around movable singularities by using
exponential or WKB-type expansions based on irregular-singular-point theory.

It has been pointed out by several authors that Eqn\rf{bureau7}
possesses two families of \Pa\ expansions
\beqn
u\approx {a_{\nu}\over z^{\nu}}
\eeqn
(where we have shifted $z-z_0$ to $z$ by using the equation's
translation invariance) distinguished by
\bes
& &\nu=2,\quad a_{2}=60\\
& &\nu=3,\quad a_{3} \ \textrm{arbitrary}
\ees
with resonances given by
\bes
& & u\approx {a_{\nu}\over z^{\nu}}+\ldots + kz^{r-\nu}\\
 & &\nu=2\  \Rightarrow\ r=-3, -2, -1, 20\\
& &\nu=3\ \Rightarrow\ r=-1, 0.
\ees
The case $\nu=2$ has a full set of resonances (even though two are negative
resonances other than the universal one). However, the case $\nu=3$ is
defective because its perturbation (in the class of \Pa\ expansions) yields no
additional degrees of freedom to the two already present in $a_3$ and
$z_0$. It is, in fact, given by the two-term expansion
\[u={a\over z^{3}}+{60\over z^2}.\]

We concentrate on this defective expansion in the remainder of this subsection.
Since this expansion allows no perturbation in the class of conventional \Pa\
expansions (which are based on regular-singular-point theory), we turn to
perturbations of the form based on irregular-singular-point theory. Consider
\[u={a\over z^{3}}+{60\over z^2}+\hat u,\]
where
\[\hat u=\exp\bigl(S(z)\bigr),\ S'\gg {1\over z}, z\ll 1.\]
(The assumption on $S'$ is to assure that $\exp(S)$ is exponential
rather than algebraic.) Substituting this into Eqn\rf{bureau7}, we get
\be
S'^4&=&{3a\over z^3}S'^2
   +\Biggl({3a\over z^3}S''+{24a\over z^4}S'-6S'^2S''\Biggr)\nn\\
   & &\quad+\Biggl({36a\over z^5}+{180\over z^2}S'^2-4S'S'''-3S''^2\Biggr)\nn\\
   & &\quad+\Biggl({960\over z^3}S'+{180\over z^2}S''-S''''\Biggr)
        + {1080\over z^4}\nn\\
   & &\quad\quad +\bigl(3S''-S'^2\bigr)e^S
     \label{bureaus}
\ee
The condition that $\exp(S)$ not be algebraic also implies that $S'^2\gg S''$,
$S'^3\gg S'''$, and $S'^4\gg S^{(IV)}$. So dividing Eqn\rf{bureaus} by $S'^2$,
taking the square root of the equation, and expanding we get
\beqn
S'={(3a)^{1/2}\over z^{3/2}}+{31\over 4z}+O\bigl(z^{-1/2}\bigr)
\eeqn
where we have used recursive substitution of the leading values of $S'$ and
$S''$ to get the term of order $1/z$.
That is, we get
\[S=-2{(3a)^{1/2}\over z^{1/2}}+{31\over 4}\log z +\textrm{const}+o(1).\]
Take one such solution, with say $a=1/3$. Then the perturbed solution
has expansion
\beqn
u = {1\over 3z^{3}}+{60\over z^2}
 +{k_{\pm} z^{31/4}}\exp\Bigl(-2/z^{1/2}\Bigr)\Bigl(1+o(1)\Bigr),
\label{expsoln}
\eeqn
where $k_{\pm}$ is an arbitrary constant. Note that there are two exponentials
here (due to the two branches of the square root of $z$) and, therefore,
$k_{\pm}$ represents
two degrees of freedom.
In the following, we consider only one of these solutions by fixing a
branch of the square root in the exponential,
say to be the one that is positive real on
the positive real semi-axis in the $z$-plane. For short, we write $k_+=k$.

Now consider the domain (or sector) of validity of this solution.
Note that the neglected terms in its expansion contain a series of
powers of $\exp(S)$ due to the
nonlinear terms in Eqn\rf{bureaus}. Therefore, for the expansion to
be asymptotically
valid, this exponential term must be bounded i.e.
\be
& &\Biggl|{k z^{31/4}}\exp\Bigl(-2/z^{1/2}\Bigr)\Biggr|<1\nn\\
& &\quad\Rightarrow |k|\exp\biggl(\Re\Bigl(-2/z^{1/2}+(31/4)\ln(z)\Bigr)\biggr)
     <1\label{inequal}
\ee
We show below that the domain of validity given by this inequality
contains a punctured disk (on a Riemann surface) whose angular width is larger
than $2\pi$.

Assume there is a branch cut along the negative semi-axis
in the $z$-plane with $\arg(z)\in (-\pi, \pi]$.
Consider $z^{1/2}$ in polar coordinates,
i.e. $z^{1/2}=re^{i\theta}$, where $-\pi/2<\theta< \pi/2$.
The positive branch will then give real part
\[\Re\Bigl(-2/z^{1/2}+(31/4)\ln(z)\Bigr)=-2\,{1\over r}\cos(\theta)
        +{31\ln r\over 4}.\]
Let $K:=\ln |k|/2$. To satisfy Eqn\rf{inequal}, we must have
\[
-\,{2\over r}\cos(\theta)
        +{31\ln r\over 4} + 2K +o(1) < 0
\]
for $r\ll 1$, i.e.
\beqn
-\cos(\theta) < -{31r\ln r\over 8} - Kr \Bigl(1 + o(1)\Bigr).
\label{polinequal}
\eeqn
Since $r$ is small (note $\ln r<0$),
this can only be violated near $\theta=\pm\pi/2$.
Fix $r$ small. Expand $\theta=\pi/2+\epsilon$. Then Eqn\rf{polinequal}
gives
\beqn
\epsilon < -\,{31r\ln r\over 8} - Kr \Bigl(1 + o(1)\Bigr)
   +O\bigl(\epsilon^3\bigr). \label{eps}
\eeqn
In particular, $\epsilon$ can be negative (so long as $|\epsilon|<1$).
A similar calculation can be made near $-\pi/2$.

These results show that the asymptotic validity of the solution
given by Eqn\rf{expsoln} can be extended to a domain which is
a disk of angular width $>2\pi$. The
small angular overlap is given by a sector of angular width $2\epsilon$
where $\epsilon$ is bounded by $O(r\ln r)$ according to Eqn\rf{eps}.

Let $z_s$ be a point in this overlapping wedge with small modulus.
At such a point, we have two asymptotic representations of $u$,
one given by a prior choice of branch of $z_s^{1/2}$ and the other
given by analytic continuation across the branch cut. If the true
solution is single-valued in this domain, the choice of two
asymptotic representations violates uniqueness. Therefore, the
true solution must itself be multivalued.
In other words, the exponential expansion shows that
Bureau's equation cannot have the \Pa\ property.

\section{Coalescence Limits}
In this section we examine asymptotic limits of integrable equations that
preserve the Painlev\'e property.  In the case of ODEs, such limits form the
basis of Painlev\'e's $\alpha$-test.  They are useful in the
identification of nonintegrable equations and may be
useful for indentifying new integrable equations
as limits of others.

Painlev\'e~\cite{painleve06} noted that under the transformation
\begin{eqnarray*}
	z &=& \E^2x-6\E^{-10},\\
	u &=& \E y +\E^{-5},\\
	\A &=& 4\E^{-15},
\end{eqnarray*}
$P_{II}$ becomes
\begin{equation}
	y''(x)=6y^2+x+\E^6\left\{ 2y^3+xy\right\}.\label{piie}
\end{equation}
In the limit as $\E$ vanishes, equation~\rf{piie} becomes $P_I$.  We write the
above limiting process as $P_{II}\longrightarrow P_I$.  Painlev\'e gave a
series of such asymptotic limits which are summarized in Figure~1.
\vskip 5mm

 \begin{figure}[ht]
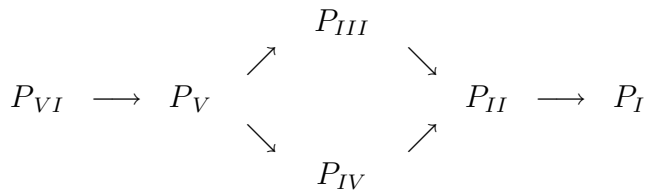

 $$
 \begin{array}{ccccccccc}
 & &     &       &\ P_{III}\       &       &       & & \\
 & &     &\nearrow &     &\searrow       &       & & \\
 P_{\,VI} & \longrightarrow & P_{\,V}
 \hskip 1mm &    &       &       &P_{II} &
 \longrightarrow & P_I\\
 & &       &\searrow &     &\nearrow       &       & & \\
 & &       &       & P_{IV}       &       &       & & \\
 \end{array}
 $$
 \caption{Asymptotic limits among the Painlev\'e equations.}
 \end{figure}

Each of these asymptotic limits coalesce the singular $u$-values of
the \Pa\ equation (see Table 2) i.e. they coalesce movable singularities.
In \cite{halburdj}, Halburd and Joshi proved that in the $P_{II}\longrightarrow
P_I$ limit,
simple poles of opposite residue coalesce to form the
double poles in solutions of $P_I$.  They also proved that
all solutions of $P_I$
can be obtained as limits of solutions of equation~\rf{piie}:

\begin{theorem}
Choose $x_0,\A,\B\in{\bf C}\/$.  Let $y_I$ and $y$ be maximally extended
solutions of $P_I$ and equation~\rf{piie} respectively, both satisfying the
initial value problem given by
$$      y(x_0)=\A,\qquad y'(x_0)=\B.    $$
Let $\Omega\subset{\bf C}$ be the domain of analyticity of $y_I\/$.
Given any compact $K\subset\Omega\/$, $\exists r_K>0$ such that $y$ is
analytic in $(x,\E)$ for $x\in K\/$ and $|\E|<r_K$. Moreover,
$y\to y_I$ in
the sup norm as $\E\to 0\/$.
\end{theorem}

The series of asymptotic (or {\em coalescence}) limits given in Figure~1
by no means represents a complete list of such limits.  The singular
$u$-values of $P_{IV}$ are $0$ and $\infty$, corresponding to zeros and poles
of the solutions respectively (the solutions of $P_{IV}$ are meromorphic
\cite{painleve02,joshik94}).  The standard coalescence limit in which
$P_{IV}$ becomes $P_{II}$ merges poles and zeros.  However, the general
solution of $P_{IV}$ contains simple poles of oppositely signed residue
which may be able to merge pairwise.
An asymptotic limit coalescing these
simple poles to form double poles does exist \cite{joshik93}.

To see this, consider a
transformation in which regions near infinity (where the poles are close
to each other) are mapped to the finite plane in the limit $\E\to 0$.
It is necessary to rescale $u$ to counter a cancellation of the oppositely
signed poles.  This leads to new variables $x$ and $w(x)$ given by
\begin{eqnarray*}
	u(z) &=& \E^pw(x),\\
	  z  &=& N+\E^qx,
\end{eqnarray*}
where $p,q>0$, $\E\ll 1$ and $N\gg 1$ is to be found in terms of $\E$.
Then $P_{IV}$ becomes
\begin{eqnarray}
w_{xx} &=& {w_x^2\over 2w} +{3\over 2}
\E^{2(p+q)}w^3+4N\E^{p+2q}w^2+8\E^{p+3q}xw^2+2(N^2-\A)\E^{2q}w \nn\\
& & \qquad\qquad+4N\E^{3q}xw+2\E^{4q}x^2w+{\B\E^{2(q-p)}\over w}.
\label{newpiv}
\end{eqnarray}
The only maximal dominant balance ({\em i.e.} a limiting state of the
equation in which a maximal number of largest terms remains~\cite{bendero})
occurs when
$$	q=p,\quad\mbox{and}\quad \A =: N^2+a\E^{-2q},	$$
where $a$ is a constant.  Then setting the largest terms $N\E^{p+2q}$ and
$N\E^{3q}$ to unity gives $N=\E^{-3p}$.  Without loss of generality redefine
$\E^p\mapsto\E$.  Then we get
$$	N=\E^{-3}\quad\mbox{and}\quad\A=\E^{-6}+a\E^{-2}.$$
Equation~\rf{newpiv} then becomes
$$
w_{xx}={w_x^2\over 2w}4w^2+(2x-a)w+{\B\over w}
+{3\over 2}\E^4w^3+8\E^4xw^2+2\E^4x^2w,
$$
or, in the limit $\E\to0$
$$
w_{xx}={w_x^2\over 2w}4w^2+(2x-a)w+{\B\over w},
$$
which is Equation~(XXXIV) (see p.340 of Ince~\cite{ince}) in the
Painlev\'e-Gambier classification of second-order differential equations
(after a simple scaling and transformation of variables).

Coalescence limits exist among PDEs also.  For example, it is
straightforward to derive the transformation~
\cite{halburdj2}
\begin{eqnarray}
        \T &=& t;\nn\\
        \xi &=& x+{3\over2\E^2}t;\nn\\
        u(x,t) &=& \E U(\xi,\T)-{1\over2\E},\nn
\end{eqnarray}
which maps the modified Korteweg-de~Vries equation (mKdV)
$$
u_t-6u^2u_x+u_{xxx}=0,
$$
to
$$      U_\T-6\E^2U^2U_\xi+6UU_\xi+U_{\xi\xi\xi}=0,     $$
which becomes the usual KdV equation in the limit $\epsilon\to0$.
An alternative method for obtaining the above asymptotic limit
is to use the $P_{II}\rightarrow P_I$ limit.  The mKdV equation is
invariant under the scaling symmetry
$$
u\mapsto \lambda^{-1}u,\quad t\mapsto \lambda^3 t, \quad x\mapsto \lambda x.
$$
Define the canonical variables
$$
z={x\over (3t)^{1/3}},\quad w=\frac13\log t,\quad u(x,t)=(3t)^{-1/3}y(z,w).
$$
In terms of these variables, the mKdV equation becomes
$$
(\underbrace{
\raisebox{-1.5ex}{ } y_{zz}-2y^3-zy-\A
\raisebox{-1.5ex}{ }}_{{P_{II}}})_z+y_w=0,
$$
where we have included the constant $\alpha$ to emphasize
its relation to $P_{II}$.
Now apply the asymptotic transformation used in the
$P_{II}\rightarrow P_I$ limit, to
determine how $y$ and $z$ transform, and transform $w$ in such a way that it
remains in the limiting form of the
equation as $\E\rightarrow 0$.  In this way
we arrive at an equation equivalent to the KdV equation,
which has a reduction to $P_I$.

In \cite{halburdj2} it is shown that the system
\begin{eqnarray}
E_x &=&\rho,\nn\\
{\widetilde E}_x &=& {\tilde\rho},\nn\\
2N_t &=& -(\rho\widetilde E+\tilde\rho E),\label{genmb}\\
\rho_t &=& NE,\nn\\
\tilde\rho_t &=& N\widetilde E,\nn
\end{eqnarray}
admits a reduction to the full $P_{III}$ ($P_{III}$ with all four constants
$\delta\ne 0$, $\A$, $\B$, $\C$ arbitrary).  We note that if $\widetilde E=
E^*$ and $\tilde\rho=\rho^*$, where a star denotes complex conjugation,
we recover the unpumped Maxwell-Bloch equations;
$$      E_x=\rho,\quad 2N_t+\rho E^*+\rho^*E=0,\quad \rho_t=NE. $$
Consider solutions of system~\rf{genmb} of the form
\begin{eqnarray*}
E       &=&     t^{-1}\varepsilon(z)w,\\
\widetilde E    &=&     t^{-1}\tilde\varepsilon(z)w^{-1},\\
N       &=&     n(z),\\
\rho    &=&     r(z)w,\\
\tilde\rho      &=& \tilde r(z) w^{-1},
\end{eqnarray*}
where $z:=\sqrt{xt}$ and $w:=(x/t)^k$, $k$ constant.  Then
$$      y(z):={\varepsilon(z)\over zr(z)}       $$
solves $P_{III}$ with constants given by
$$
\A=2(r\tilde\varepsilon-\tilde r\varepsilon+4k),
\quad\B=4(1+2k),\quad\C=4(n^2+r\tilde r),\quad\D=-4.
$$
Note that by rescaling $y$ we can make $\delta$ any nonzero number.

Using the procedure outlined for mKdV $\rightarrow$ KdV, it can be shown
\cite{halburdj2} that the $P_{III}\rightarrow P_{II}$ limit
induces an asymptotic limit in which the generalized unpumped Maxwell-Bloch
system~\rf{genmb} becomes the dispersive water-wave equation (DWW)
$$
u_{xxxx}+2u_tu_{xx}+4u_xu_{xt}+6u_x^2u_{xx}+u_{tt}=0,
$$
which is known to admit a reduction to $P_{II}$ (Ludlow and
Clarkson~\cite{ludlowc}).
The $P_{II}\rightarrow P_I$ limit then gives DWW $\rightarrow$ KdV.
Ludlow and Clarkson~\cite{ludlowc} have shown that DWW also admits a symmetry
reduction to the full $P_{IV}$.  The $P_{IV}
\rightarrow P_{II}$ limit then
induces an asymptotic limit that maps DWW back to
itself in a nontrivial way.  The limit $P_{IV}\rightarrow P_{34}$,
outlined above, induces another limit in which DWW
is mapped to the KdV.
All six Painlev\'e equations are known to arise as reductions of the self-dual
Yang-Mills (SDYM)
equations (Mason and Woddhouse \cite{masonw}).  Asymptotic limits
between the Painlev\'e equations can be used to induce similar limits
within the SDYM system (Halburd~\cite{halburd}).

\section{Acknowledgements}
The work reported in this paper was partially supported by the Australian
Research Council. The authors also gratefully acknowledge with thanks the
efforts
of the organizing committee, particularly Dr K. ~M. Tamizhmani, and CIMPA in
arranging the winter school on Integrable Systems at Pondicherry.

\end{document}